\documentclass[submission, Proceedings]{SciPost}

\usepackage{epsfig}
\usepackage{multirow}
\usepackage{multicol}
\usepackage{hyperref}
\usepackage{slashed}
\usepackage{amssymb}
\usepackage{amsmath}
\usepackage{amsfonts}
\usepackage{diagbox}

\newcommand{\be}{\begin{equation}}
\newcommand{\bfr}{\begin{frame}}
\newcommand{\ee}{\end{equation}}
\newcommand{\efr}{\end{frame}}
\newcommand{\nn}{\nonumber}
\newcommand{\pa}{\partial}
\newcommand{\ba}{\begin{eqnarray}}
\newcommand{\ea}{\end{eqnarray}}

\newcommand{\unit}{\mathbb{I}}
\newcommand{\sumint}{\sum_n \int}

\newcommand{\blue}{\textcolor{blue} }
\newcommand{\red}{\textcolor{red} }

\hypersetup{
    colorlinks,
    linkcolor={red!50!black},
    citecolor={blue!50!black},
    urlcolor={blue!80!black}
}

\usepackage[bitstream-charter]{mathdesign}
\DeclareSymbolFont{usualmathcal}{OMS}{cmsy}{m}{n}
\DeclareSymbolFontAlphabet{\mathcal}{usualmathcal}

\begin{document}

\begin{center}{\Large \textbf{
Chiral symmetry restoration with three chiral partners\\
}}\end{center}

\begin{center}
J. M. Torres-Rincon\textsuperscript{1,2$\star$}
\end{center}

\begin{center}
{\bf 1} Departament de F\'isica Qu\`antica i Astrof\'isica and Institut de Ci\`encies del Cosmos (ICCUB), Facultat de F\'isica,  Universitat de Barcelona, Mart\'i i Franqu\`es 1, 08028 Barcelona, Spain
\\
{\bf 2} Institut f\"ur Theoretische Physik, Goethe Universit\"at Frankfurt, Max von Laue Strasse 1, 60438 Frankfurt, Germany
\\
*torres-rincon@ub.edu
\end{center}

\begin{center}
\today
\end{center}


\definecolor{palegray}{gray}{0.95}
\begin{center}
\colorbox{palegray}{
  \begin{minipage}{0.95\textwidth}
    \begin{center}
    {\it  XXXIII International (ONLINE) Workshop on High Energy Physics \\“Hard Problems of Hadron Physics:  Non-Perturbative QCD \& Related Quests”}\\
    {\it November 8-12, 2021} \\
    \doi{10.21468/SciPostPhysProc.?}\\
    \end{center}
  \end{minipage}
}
\end{center}

\section*{Abstract}
{\bf
I discuss the masses of chiral partners in the context of chiral symmetry restoration at finite temperature. Using the Nambu--Jona-Lasinio model I first remind the usual situation where two mesons of opposed parity become degenerate above the chiral transition temperature. Then I consider an effective theory for $D$ mesons where the positive parity companion presents a ``double pole structure''. In this case three different masses need to be analyzed as functions of the temperature. I suggest a possible restoration pattern at high temperatures when the back-reaction of the quark condensate is incorporated.
}

\vspace{10pt}
\noindent\rule{\textwidth}{1pt}
\tableofcontents\thispagestyle{fancy}
\noindent\rule{\textwidth}{1pt}
\vspace{10pt}

\section{Introduction}
\label{sec:intro}

For $N_f$ flavors of quarks with nearly zero current masses, the QCD Lagrangian presents an approximate chiral symmetry $SU_L(N_f) \times SU_R(N_f)$, which is spontaneously broken in vacuum down to $SU_V(N_f)$. Then, the effective masses of low-energy excitations with negative and positive parities ($\pi$ and $\sigma/f_0(500)$, respectively) become splitted. Nevertheless it is accepted that the presence of a medium can restore the chiral symmetry at high temperatures or densities following a phase transition, which can be of first-, second-order or a crossover.

In this contribution I focus on chiral symmetry restoration at finite temperature, reflected in a partial degeneration of chiral partner masses for $T>T_c$, where $T_c$ is the transition temperature. In QCD with physical quarks, the chiral transition at vanishing net-baryon density is known to be an analytic transition~\cite{Aoki:2006we}. Different lattice-QCD calculations at finite temperature show this degeneracy for the screening masses of mesonic excitations with different spins~\cite{Detar:1987kae,Bazavov:2019www}.

In Ref.~\cite{Torres-Rincon:2021wyd} I classified different effective field theories (EFTs) of QCD according to the nature of the chiral partners. These could be either fundamental degrees of freedom---represented by quantum fields in the effective action---or generated dynamically via a few-body (e.g. Bethe-Salpeter) equation. In both cases the temperature modifies the propagation of the particles incorporating corrections to their masses. In the former case the fundamental states acquire quasiparticle properties in the medium, while in the second case the emergent collective excitations have a temperature-dependent pole mass, due to thermal corrections in the few-body equation. The classification introduced in Ref.~\cite{Torres-Rincon:2021wyd} is summarized in Table~\ref{tab:class} for the $\pi/\sigma$ system.

\begin{table}[ht]
\begin{tabular}{|c|c|c|}
\hline
\diagbox{$J^P=0^-$}{$J^P=0^+$}  &  Fundamental d.o.f. & Dynamical d.o.f.\\
\hline
\multirow{4}{*}{Fundamental d.o.f.}& {\bf Linear $\sigma$ model} &  \\
     & \cite{Coleman:1974jh},\cite{Bochkarev:1995gi},\cite{Dobado:2009ek}  & {\bf Chiral perturbation theory}\\
     & {\bf Quark-meson model}  & \cite{Dobado:2002xf},\cite{GomezNicola:2002an} \\
     & \cite{Jungnickel:1995fp},\cite{Scavenius:2000qd} &  \\
\hline
\multirow{4}{*}{Dynamical d.o.f.}     &   \multirow{4}{*}{-}         & {\bf Nambu--Jona-Lasinio model}   \\ 
    &            &  \cite{Vogl:1991qt},\cite{Klevansky:1992qe} \\
      &            & {\bf Polyakov--NJL model}   \\ 
    &            &  \cite{Ratti:2005jh},\cite{Torres-Rincon:2015rma}\\ 
\hline
\end{tabular}
\centering
\caption{Different models/effective theories where the chiral partners $\pi (J^P=0^-)$ and $\sigma (J^P=0^+)$ can be both fundamental degrees of freedom, dynamically generated states, or a hybrid scenario with one fundamental and one emergent state.}\label{tab:class}
\end{table}

I first review the case where both chiral companions are generated dynamically. Thus the $\pi$ and $\sigma$ are described by collective mesonic excitations emerging out of the quark-antiquark attractive interaction. For this goal I focus on the Nambu--Jona-Lasinio (NJL) model of interacting massive quarks in Sec.~\ref{sec:pnjl}.    Then I consider a novel case where one of the chiral partners is a dynamically generated state, but presents a ``two-pole structure''. This is exemplified by the positive parity $D_0^*(2300)$ resonance (chiral partner of the $D$ meson). As a function of the temperature, the evolution of its pole mass was considered in Refs.~\cite{Montana:2020lfi,Montana:2020vjg}. This case is discussed in Sec.~\ref{sec:chiral}, where I conjecture on the possible chiral restoration pattern for three states when the transition temperature is approached. Conclusions are given in Sec.~\ref{sec:conclu}.

\section{Two chiral partners: Nambu–Jona-Lasinio model} \label{sec:pnjl}

I start by reviewing the common situation with two chiral companions. In particular I cover the case where both states are emerging out of the Bethe-Salpeter equation for the quark-antiquark scattering. For this goal I use the NJL model~\cite{Nambu:1961tp,Vogl:1991qt} which describes the low-energy interaction of quarks and antiquarks. 
  
The minimal NJL Lagrangian contains two flavors ($u$ and $d$) of quarks interacting locally in both scalar and pseudoscalar spin channels with the same coupling $\blue{{\cal G}}$,
\begin{align}
 {\cal L}_{\textrm{NJL}} &= \sum_{l=u,d} \bar{\psi}_l (i \slashed{\pa}-m_{0l}) \psi_l \nn \\
&+ \blue{{\cal G}} \sum_{a} \sum_{ijkl} \left[ (\bar{\psi}_i \ i\gamma_5 \ \tau^{a}_{ij} \psi_j) \ (\bar{\psi}_k \ i \gamma_5 \ \tau^{a}_{kl} \psi_l)
+ (\bar{\psi}_i \ \unit \ \tau^{a}_{ij} \psi_j) \ 
(\bar{\psi}_k  \ \unit \  \tau^{a}_{kl} \psi_l) \right]  , \label{eq:lagPNJL}
 \end{align}
where the quark field is labeled with flavor indices $i,j,k,l=\{u,d\}$, $\tau^{a}$ ($a=1,2,3$) are the flavor generators of $SU_f(2)$ algebra, and $\unit$ is the 4x4 unit matrix of Dirac space.

The bare quark masses $m_{0l}$ are dressed by the effects of interactions. At mean-field level these are given by the quark condensate~\cite{Buballa:2003qv,Torres-Rincon:2015rma}, 
\be m_i = m_{i0} - 4 \blue{{\cal G}} \langle \langle \bar{\psi}_i \psi_i \rangle \rangle \ . \label{eq:mass} \ee

This condensate acts as an order parameter of the chiral transition, when considered a function of the temperature. Using the imaginary-time formalism it reads,
\be \langle \langle \bar{\psi}_i \psi_i \rangle \rangle = N_c \ \textrm{tr}_\gamma \sumint_q \frac{1}{\slashed{q}-m_i} \ , \ee 
where $N_c=3$, tr$_\gamma$ denotes the trace in Dirac space, and $\sumint_q$ is the Matsubara summation followed by the 3-momentum integration. Notice that $q^\mu=(i\omega_n,{\bf q})$.

\begin{figure}[h]
\begin{center}
\includegraphics[width=0.4\textwidth]{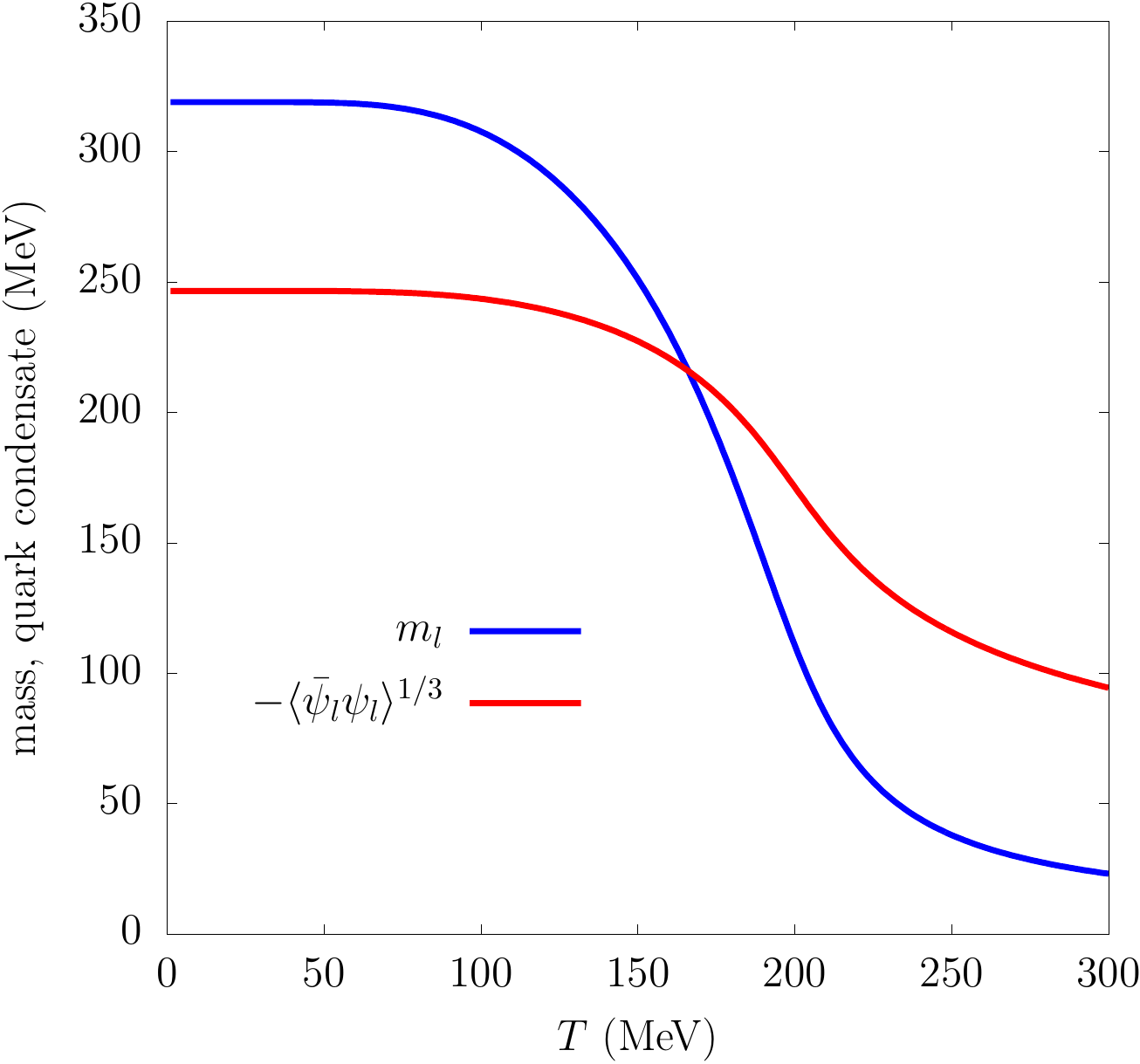}
\hspace{15mm}
\includegraphics[width=0.48\textwidth]{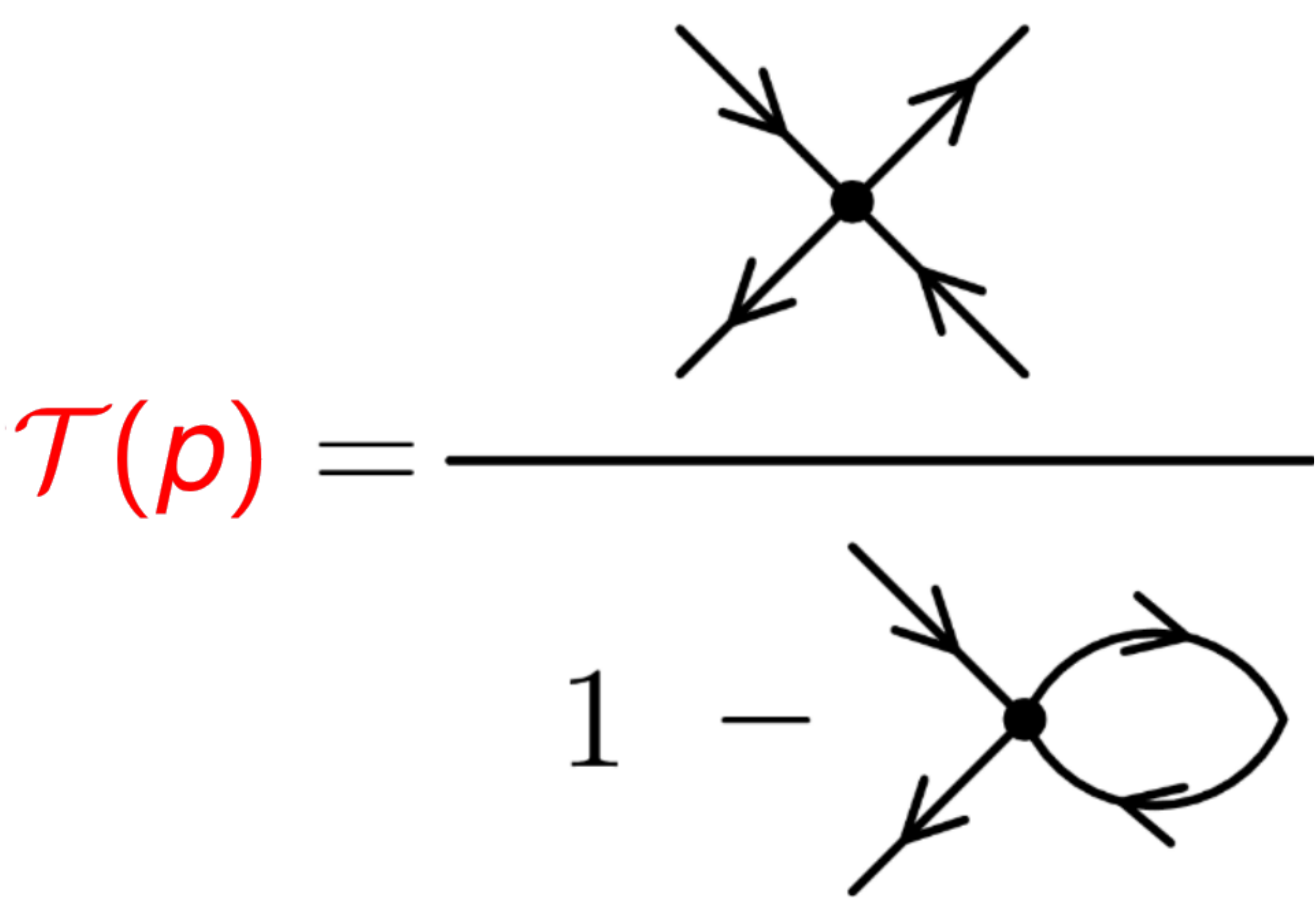}
\caption{Left: Quark mass (blue) and quark condensate (red) as functions of the temperature of the 2-flavor NJL model within the mean-field approximation. Right: Diagrammatic solution of the $T$-matrix equation~(\ref{Tmatrix}) in terms of the quark-antiquark interaction vertex of the NJL model $\blue{{\cal G}}$ (represented by the black circles). \label{fig:tran}}
\end{center}
\end{figure}

The transition to the chirally-restored phase at high temperature is not only signalled by the quark condensate but also by the quark mass~(\ref{eq:mass}). Both are shown in the left panel of Fig.~\ref{fig:tran} as functions of the temperature. I used the parameter set of the $N_f=2$ NJL model given in~\cite{Blaschke:2013zaa}.

Mesonic states can be generated after solving the Bethe-Salpeter equation for the quark-antiquark scattering. It is given within the random-phase approximation using the imaginary time formalism~\cite{Vogl:1991qt,Torres-Rincon:2015rma}. Suppressing unnecessary flavor indices, the equation reads,
\be \red{{\cal T} (p)} = \blue{{\cal G}} + \blue{{\cal G}} \ \Pi (p) \ \red{{\cal T} (p)}  \ , \label{eq:BSPNJL}
\ee
where $p=(i \nu_m,{\bf p})$, and the quark-antiquark propagator is also calculated at finite temperature as,

\be \label{eq:polmeson} \Pi (i\nu_m,{\bf p}) = - \sumint_k \textrm{ tr}_{\gamma}
 \left[ \bar{\Omega} \ S \left(i\omega_n , {\bf k} \right) \ \Omega \ S \left( i\omega_n-i\nu_m ,{\bf k} - {\bf p} \right)  \right] \ ,
 \ee
where $S$ denote (anti)quark dressed propagators, and $\Omega$ contains (apart from flavor and color factors) the two possible Dirac structures $i\gamma_5$ and $\unit$, needed to generate pseudoscalar and scalar meson excitations. The quark-antiquark propagator
provides a non-trivial analytical structure to the final scattering amplitude $\red{{\cal T} (p)}$. The solution of the two-body equation~(\ref{eq:BSPNJL}) reads
\be \label{Tmatrix} \red{{\cal T} (p)}  =  \frac{ \blue{{\cal G}} }{1- \blue{{\cal G}}  \ \Pi(p) } \ , \ee
where the external Matsubara frequency $i\nu_m$ is eventually extended to the entire complex energy plane $z$. A diagrammatic version of this equation is given in the right panel of Fig.~\ref{fig:tran}. Notice that the denominator can accommodate poles in different regions of the complex plane signalling the generation of bound and scattering states. 

\begin{figure}[h]
\begin{center}
\includegraphics[width=0.32\textwidth]{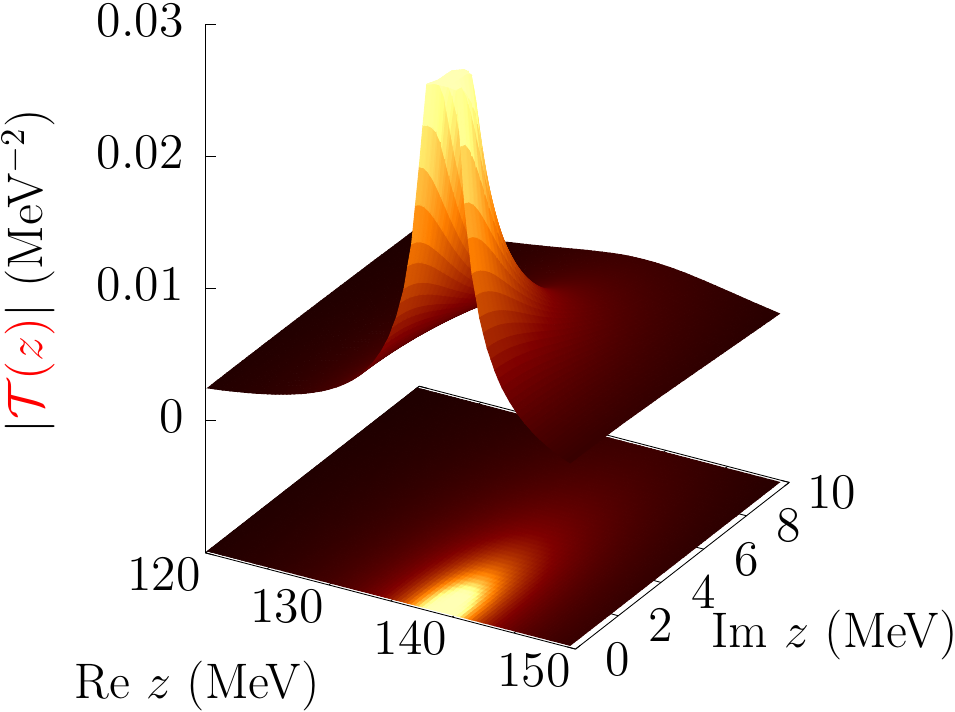}
\includegraphics[width=0.32\textwidth]{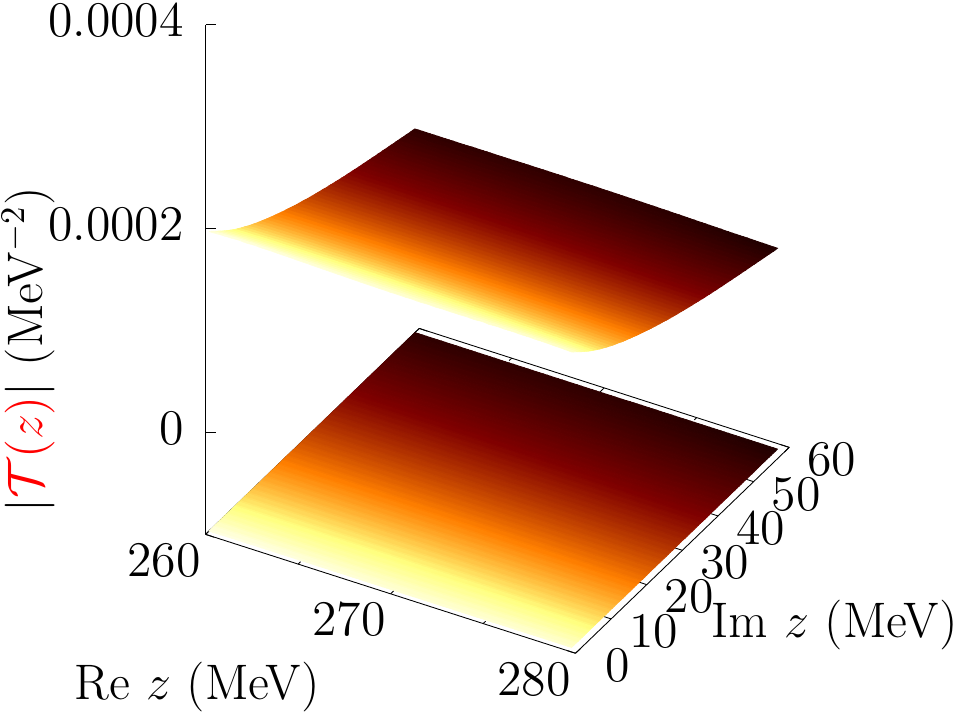}
\includegraphics[width=0.32\textwidth]{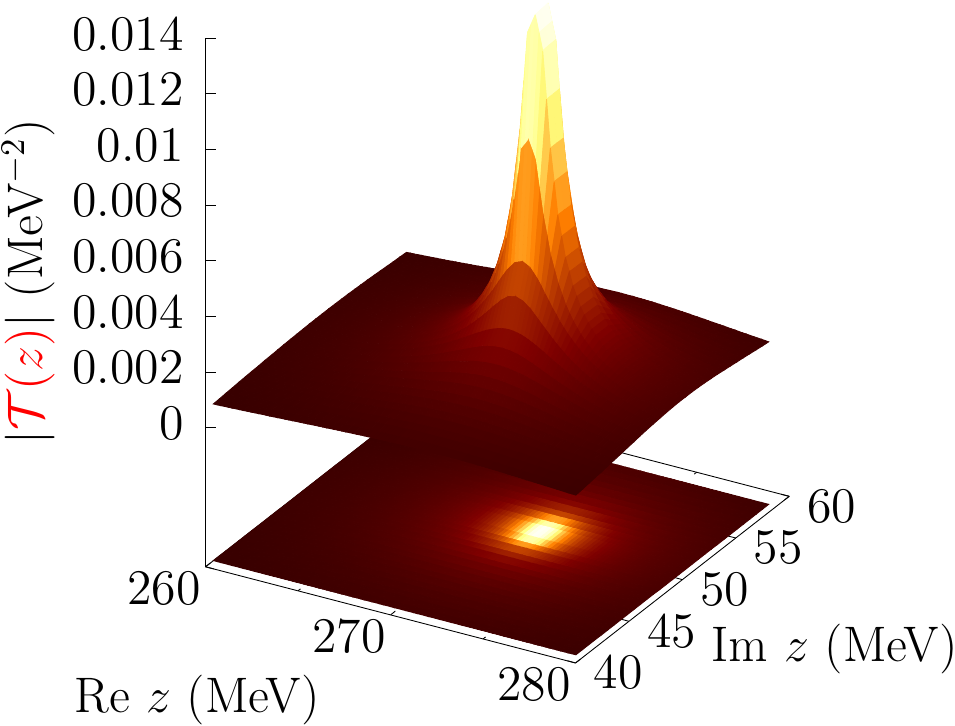}
\caption{Left: $\pi$ pole in the complex energy plane of the first Riemann surface at $T=25$ MeV. Middle: No pole in the complex energy plane of the first Riemann surface at $T=250$ MeV. Right: $\pi$ pole after analytic continuation to the second Riemann surface at $T=250$ MeV.\label{fig:pion}}
\end{center}
\end{figure}

For example, in the $J^P=0^-$ channel in vacuum and at low temperatures, the pion excitation emerges in the real axis as a quark-antiquark bound state. This is seen in the left panel of Fig.~\ref{fig:pion} for $T=25$ MeV, where the first Riemann surface of $\red{{\cal T} (p)}$ is plotted.

\vspace{5mm}

At high temperatures the quark mass decreases, see Fig.~\ref{fig:tran}, and the pion mass starts to increase becoming a resonant state with a thermal decay width. It is natural that no such solution can be obtained without an analytic continuation to the second Riemann sheet of $\red{{\cal T}(p)}$ (see middle panel of Fig.~\ref{fig:pion}). Previous studies in the NJL model avoided this analytic continuation by introducing some approximations or assuming a quasiparticle picture even when the decay width can become of the same order as the pion mass. However it is not difficult to perform the required analytic continuation of the quark-antiquark propagator above the two-quark mass threshold. For a complex value of the energy $z$,
\be \Pi^{II} (z,{\bf p};T) = \Pi^{I} (z,{\bf p};T) - 2 i \textrm{Im} \Pi^{I} (z,{\bf p};T) \qquad \textrm{for Re } z>2m_q(T) \ . \ee

After this analytic continuation, the $\red{ {\cal T}}$ matrix in the second Riemann surface acquires a pole with finite imaginary part as seen in the right panel of Fig.~\ref{fig:pion}. The pion generated as $T=250$ MeV is interpreted as a resonant state.

\begin{figure}[h]
\begin{center}
\includegraphics[width=0.4\textwidth]{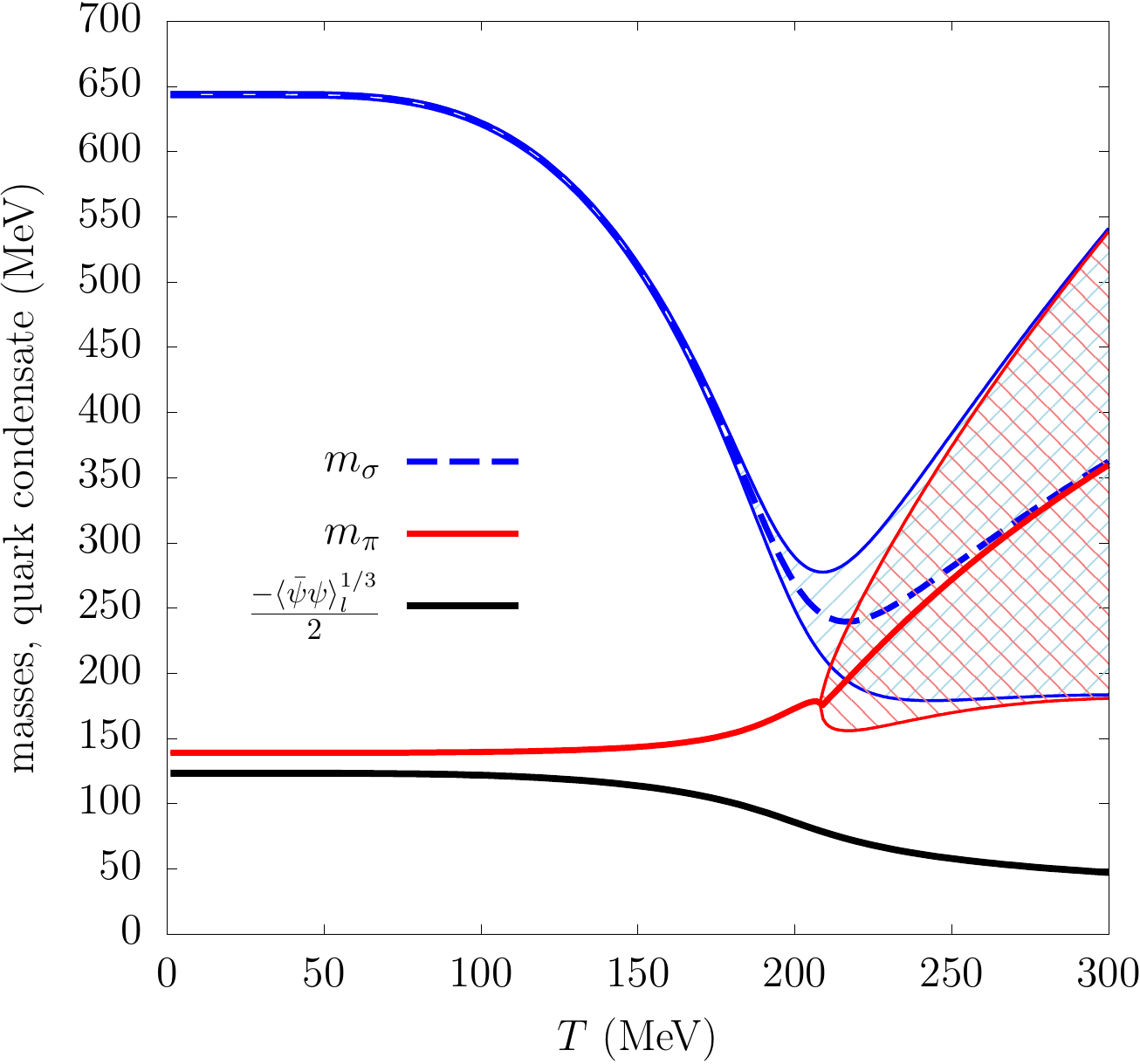}
\caption{$\pi$ (red solid line) and $\sigma$ (blue dashed line) masses as functions of the temperature, below and above the chiral phase transition (signalled by the quark condensate in black line). The decay width is plotted as bands around the masses.\label{fig:scalars}}
\end{center}
\end{figure}

A plot of the $\pi$ and $\sigma$ masses (real parts of their pole positions) as functions of the temperature is given in Fig.~\ref{fig:scalars}. Their large splitting at low temperatures vanishes around $T=250$ MeV, where chiral symmetry is effective restored. The imaginary part of the poles---interpreted as thermal decay half width---is represented as a band around the masses. This decay probability into quarks become also equal in both sectors at high temperatures, pointing to a full degeneracy of the spectral shapes in this region.  Chiral symmetry is restored for states not originally present in the effective Lagrangian. 

\section{Three chiral companions: Covariant chiral EFT} \label{sec:chiral}

Now I turn to a case belonging to the class of the upper right corner of Table~\ref{tab:class}. Here the negative parity state is part of the initial degrees of freedom, while the positive parity companion is generated via a two-body equation. The novel feature is that the emergent state is not identified with a single pole of the scattering amplitude, but it consists of a ``two-pole structure''. The example is taken from charmed mesons at finite temperature, as presented in Refs.~\cite{Montana:2020lfi,Montana:2020vjg}. Three chiral states are found in the nonstrange sector $S=0$ corresponding to the $D$ and $D_0^*(2300)$ partners, while in the strange sector $S=1$ one finds the usual parity doublet with the $D_s$ and $D_{s0}^*(2317)$ mesons.

Open charm mesons can be studied with an EFT approach implementing a combination of chiral and heavy-quark symmetry in a covariant way~\cite{Kolomeitsev:2003ac,Guo:2008gp,Geng:2010vw,Abreu:2011ic}. Details of the effective Lagrangian can be found in the recent~\cite{Montana:2020vjg} (and references therein), where both heavy ($D,D_s$) and light $(\pi,K,{\bar K},\eta)$ ground states are the fundamental degrees of freedom. The EFT provides the tree-level amplitudes for heavy-light meson scattering. At leading order~\cite{Guo:2009ct,Geng:2010vw} (see also~\cite{Montana:2020lfi,Montana:2020vjg} for next-to-leading order results),
\be \blue{V (k,k_3 \rightarrow k_1,k_2)} = \frac{C_0}{4f_\pi^2} \ \left[ (k+k_3)^2 - (k-k_2)^2 \right] \ , \label{eq:V} \ee
where $C_0$ are known isospin coefficients, and $f_\pi$ is the pion decay constant. Eq.~(\ref{eq:V}) includes all possible elastic and inelastic channels (given by the corresponding $C_0$).

The $s-$wave projected amplitudes from~(\ref{eq:V}) can be incorporated into a Bethe-Salpeter equation---similar to the one in the NJL model in Eq.~(\ref{eq:BSPNJL})---to calculate the resumed (and unitary) scattering amplitudes $\red{ {\cal T}}$. Applying the ``on-shell factorization'' method~\cite{Oller:1997ti} to simplify the equation one gets,
\be \label{eq:BetheSalpeter} \red{{\cal T} (s)}= \blue{V (s)} + \blue{V(s)} \ G_2(s) \ \red{{\cal T}} \ (s) \ , \ee
where the two-meson propagator $G_2(s)$ plays the same role of the quark-antiquark propagator in the NJL model~(\ref{eq:polmeson}). The formal solution is also similar to Eq.~(\ref{Tmatrix}),
\be \red{{\cal T} (s)}=  \frac{ \blue{V (s)}}{ 1- \blue{V(s)} G_2 (s) } \ , \label{eq:Tmatrix2} \ee
but in this case it is fully given in a coupled-channel approach (see details in~\cite{Montana:2020lfi,Montana:2020vjg}).

\begin{figure}[h]
\begin{center}
\includegraphics[width=0.32\textwidth]{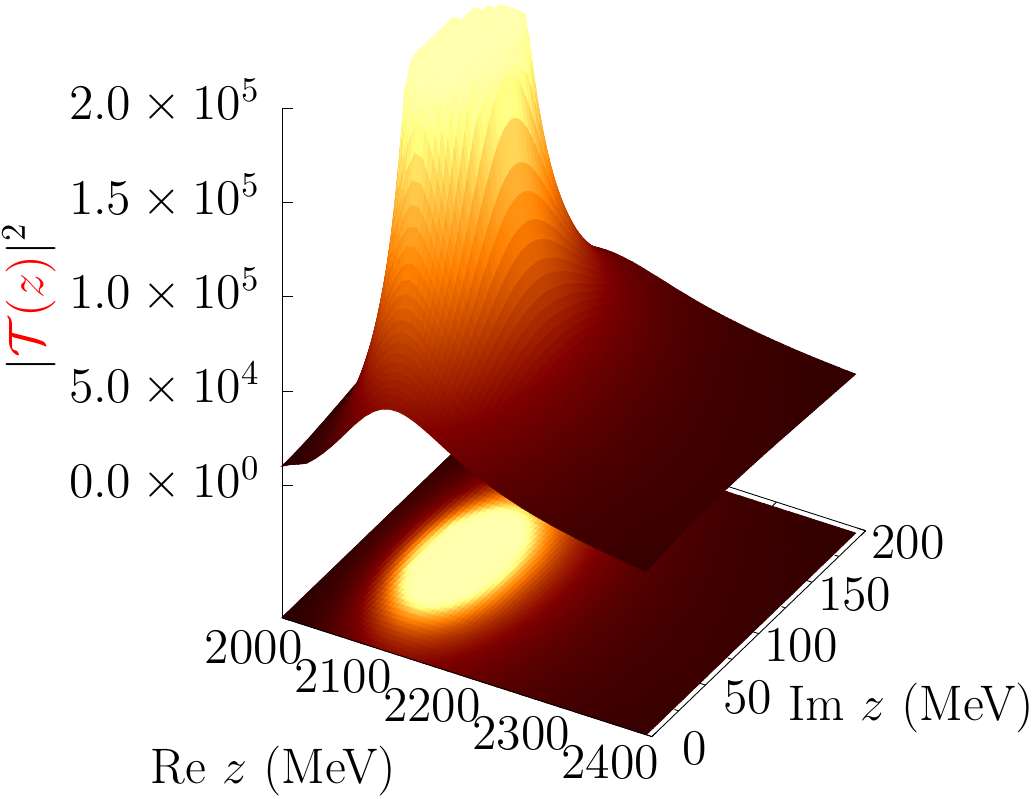}
\includegraphics[width=0.32\textwidth]{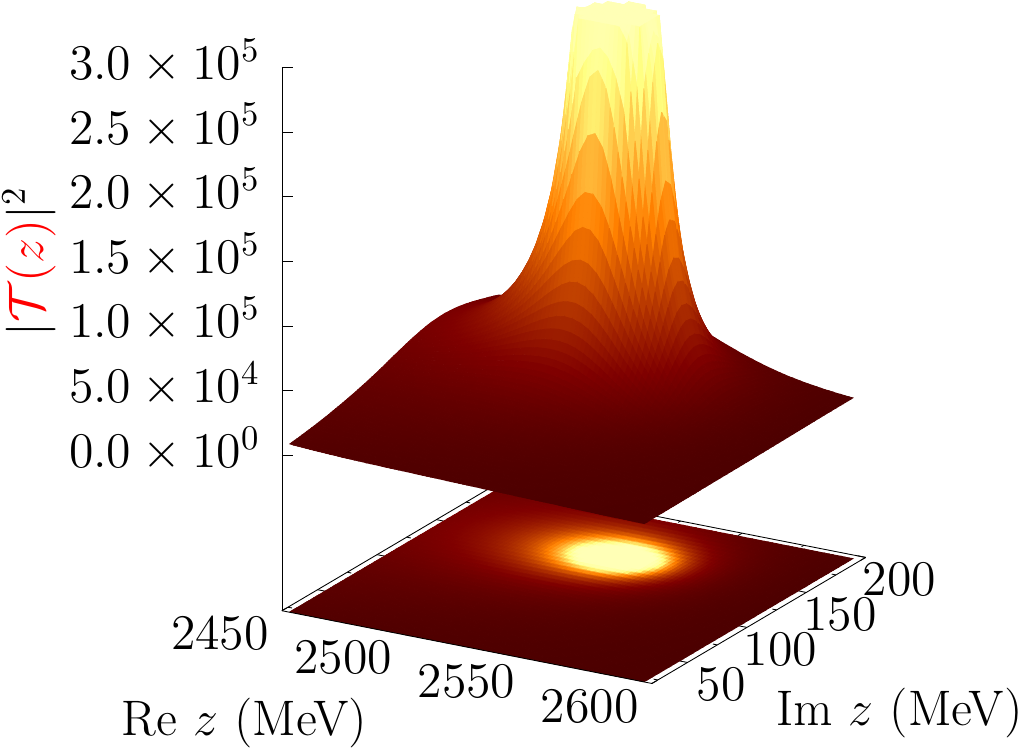}
\includegraphics[width=0.32\textwidth]{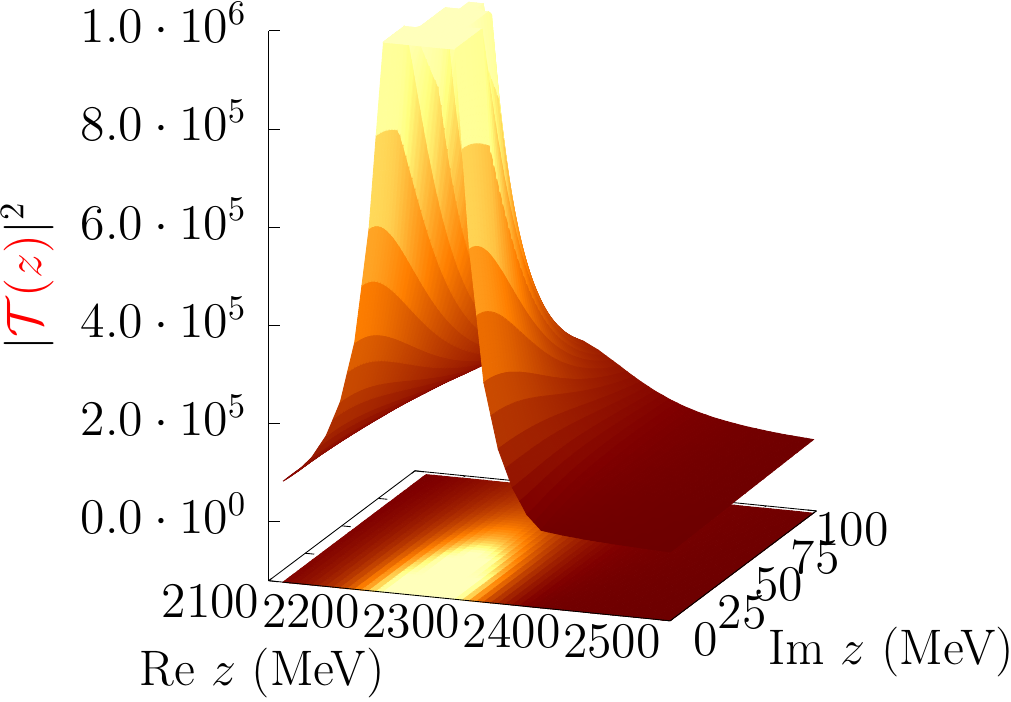}
\caption{Generated poles at $T=0$ in the heavy-light meson system. Left: Lower pole of the $D_0^*(2300)$ resonance. Middle: Higher pole of the $D_0^*(2300)$ resonance. Right: Pole of the $D_{s0}^*(2317)$ bound state. Figure taken from~\cite{Torres-Rincon:2021wyd}. \label{fig:charm}}
\end{center}
\end{figure}

Looking for poles of Eq.~(\ref{eq:Tmatrix2}) in the complex energy plane of the different channels, one finds the dynamically-generated states. In vacuum, the poles in the $S=0$ channel are shown in the left and middle panels of Fig.~\ref{fig:charm} and correspond to the $D_0^*(2300)$, the chiral partner of the $D$ meson. In the $S=1$ channel one finds the pole on the right panel of Fig.\ref{fig:charm} which is identified with the very narrow $D_{s0}^*(2317)$, the chiral companion of the $D_s$.

Both poles in the left and middle panels correspond to the physical resonance $D_0^*(2300)$. It is known that this state consists of a ``two-pole structure''~\cite{Albaladejo:2016lbb,Meissner:2020khl}, where both poles share the same quantum numbers and can interfere at the real energy (physical) line. Nonetheless they couple with different strength to the possible decay channels~\cite{Montana:2020vjg}.  

Focusing on the masses of these states (together with those of the ground states, $D$ and $D_s$) one can extend the $T$-matrix equation (\ref{eq:Tmatrix2}) to finite temperature. This has been performed in Refs.~\cite{Montana:2020lfi} using the imaginary time formalism. It is important to mention that self-consistency is required at finite temperature as the thermal corrections to the meson propagators need to be introduced in the $T$-matrix equation. 

\begin{figure}[h]
\begin{center}
\includegraphics[width=0.47\textwidth]{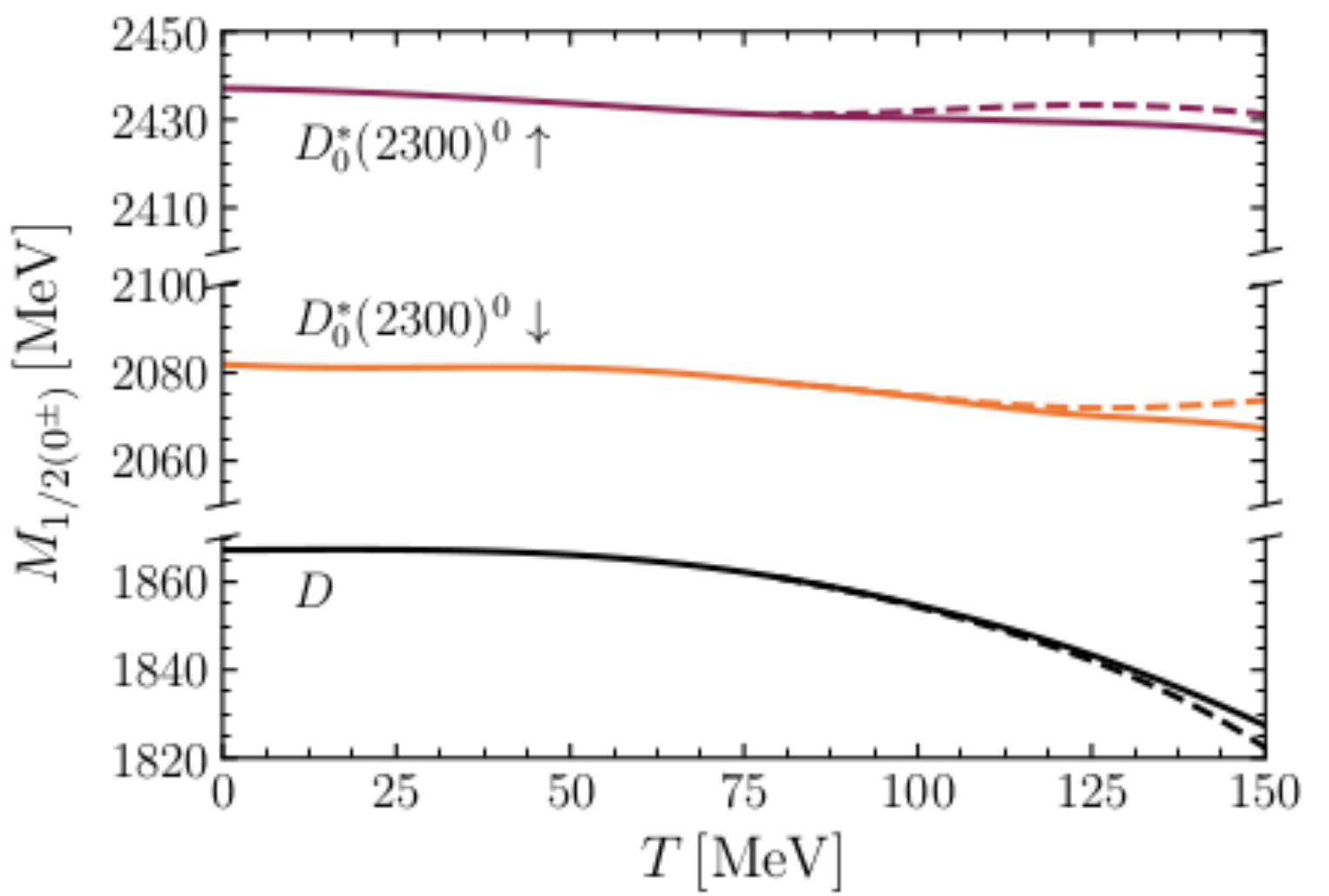}
\hspace{5mm}
\includegraphics[width=0.47\textwidth]{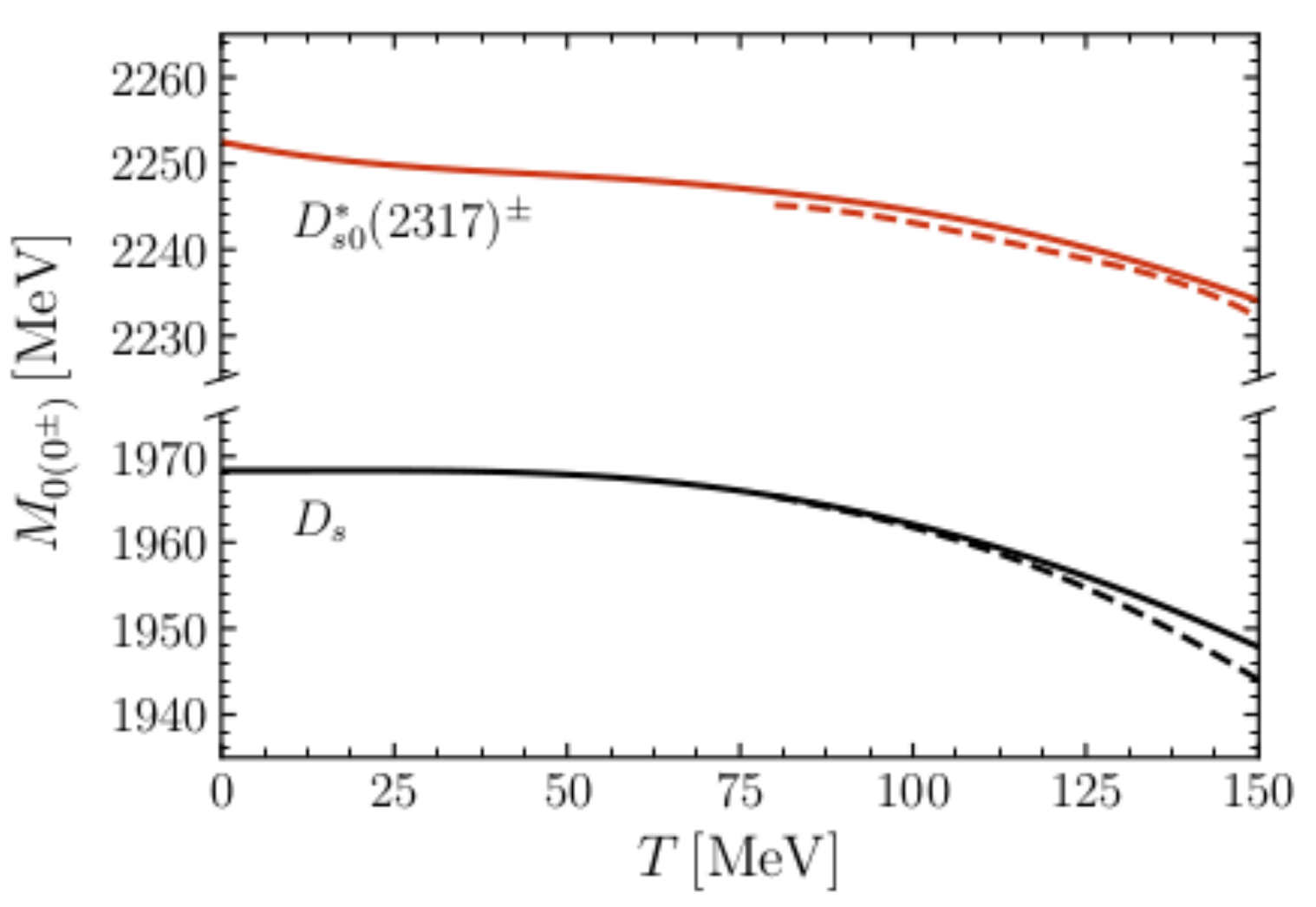}
\caption{\label{fig:thermalDJ0} Thermal masses of chiral partners in the $D$-meson sector. Left: $S=0$ channel with the $D$ meson and the $D_0^*(2300)$ resonance (double pole). Right: $S=1$ channel with the $D_s$ meson and the $D_{s0}^* (2317)$ bound state. Figure taken from~\cite{Montana:2020lfi}.}
\end{center}
\end{figure}

The resulting thermal masses are given in Fig.~\ref{fig:thermalDJ0} as functions of the temperature. The left panel shows the $S=0$ case with the ground state and the two poles of the $D_0^*(2300)$ resonance. The right plot contains the $S=1$ case with the $D_s$ and the $D_{s0}^*(2317)$. Solid lines in Fig.~\ref{fig:thermalDJ0} represent the results containing the only effect of the pion in the heavy-meson dressing, while the dashed lines account for the additional $K,{\bar K}$ contribution. Due to the Boltzmann suppression the kaonic contribution is very small even at $T=150$ MeV. 

There is a general mass dropping with temperature which, in relative terms, is very small ca. $\sim 2 \%$ of their vacuum values. Given such a tiny mass reduction even at the highest temperature, and the fact the EFT cannot be applicable beyond it~\cite{Montana:2020lfi}, one concludes that no mass degeneracy is expected from the accounting of the thermal effects alone. It is also not possible to conclude the precise pattern between the two poles of the $D_0^*(2300)$ resonance, as both follow in parallel a reduction of $\Delta m \sim -10$ MeV at $T=150$ MeV. In Ref.~\cite{Montana:2020vjg} we studied the case where the pion mass receives a thermal correction---due to interactions with other light mesons---but no significant different picture was obtained.

From the study of the NJL model (but also other EFTs) it is clear that the effect of the thermal chiral condensate---whose decrease manifests the approach to the chiral transition---is a key ingredient to account for the mass degeneracy. In the worked approximation such an effect does not appear, and the model does not know about any phase transition at high temperature. 

The reduction of the chiral condensate seen in lattice-QCD calculations should affect the pseudo-Goldstone bosons properties. At low temperatures, where chiral perturbation theory is applicable, this effect can be seen at the level of the Gell-Mann--Oakes--Renner relation at finite temperature~\cite{Gasser:1986vb,Pisarski:1996mt}. In particular, $f_\pi(T)$ acquires a reduction when the temperature is increased toward the transition temperature~\cite{Gasser:1986vb,Pisarski:1996mt},
\be \frac{f_\pi(T)}{f_{\pi,0}} \simeq 1- \frac{T^2}{12f_{\pi,0}^2} \ . \ee
according to chiral perturbation theory at leading order (and also linear sigma model~\cite{Bochkarev:1995gi}).

As a simple calculation to gauge the dependence of $f_\pi$ on the charmed mesons masses, I apply a reduction to this parameter from its vacuum value to the vacuum calculation of the $T$-matrix equation~(\ref{eq:Tmatrix2})~\cite{Torres-Rincon:2021wyd}. This isolates the effect of $f_\pi$, as pure thermal effects played a small role in the meson masses (cf. Fig~\ref{fig:thermalDJ0}).

\begin{figure}[h]
\begin{center}
\includegraphics[scale=0.55]{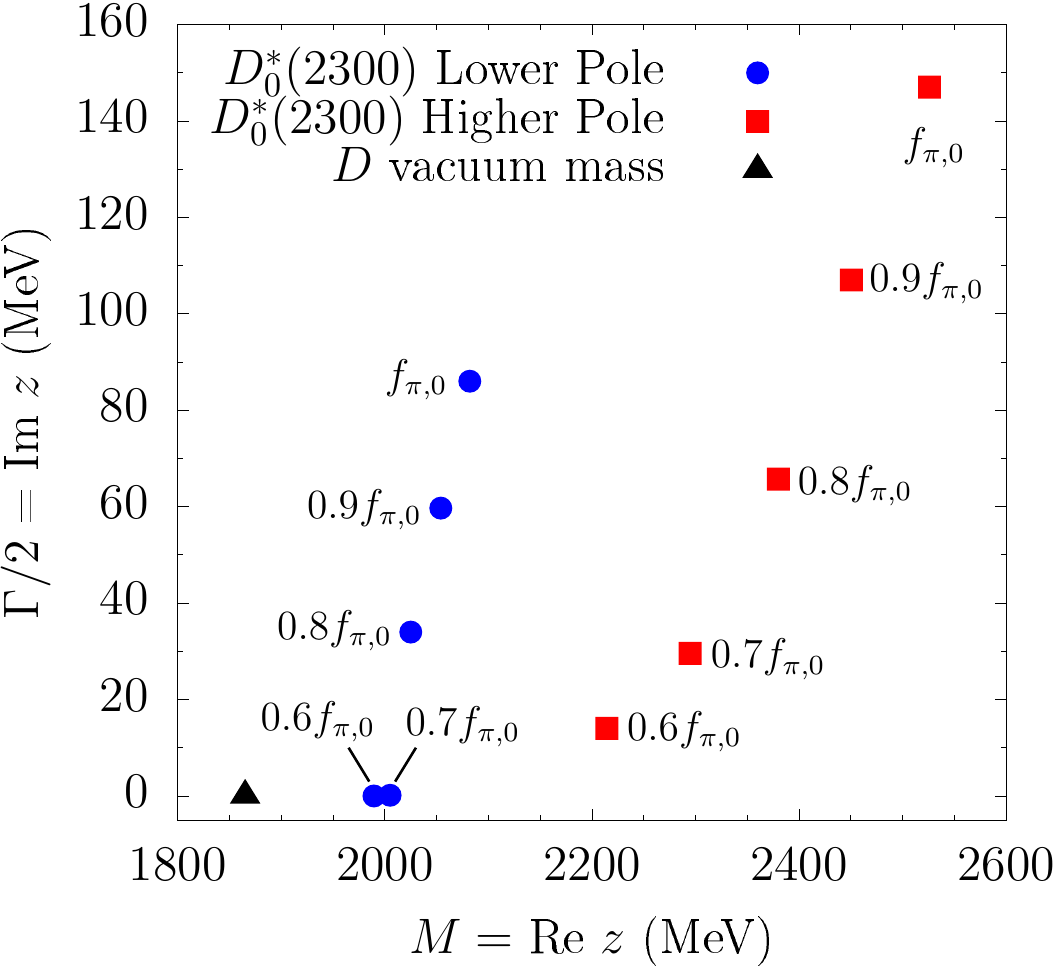}
\includegraphics[scale=0.55]{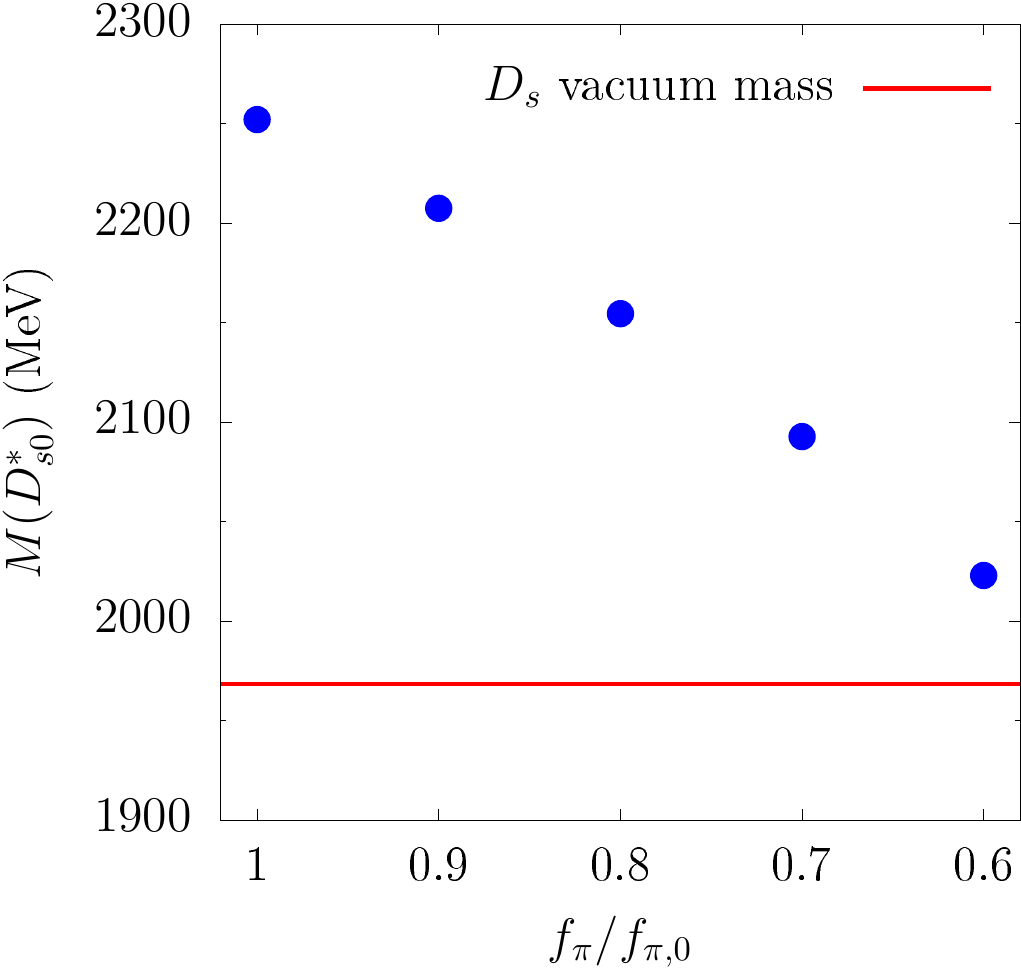}
\caption{Pole positions of the meson charm states as functions of $f_pi$, when it is reduced with respect to its vacuum value. Left: $(J,S)=(0,0)$ channel where the ``two-pole structure'' of the  $D_0^*(2300)$ is generated. Right: $(J,S)=(0,1)$ sector where the bound state $D_{s0}^* (2317)$ appears.\label{fig:test}}
\end{center}
\end{figure}

In the right panel of Fig.~\ref{fig:test} I present the masses of the two chiral partners in the $S=1$ channel as a function of $f_\pi/f_\pi(T=0)$. The (input) mass of the $D_s$ is kept fixed, but the modification of $f_\pi$ produces a bound state with a sizable mass reduction approaching closely to its chiral partner, as expected. In the left plot I show the location of the poles in the complex energy plane for the $S=0$ case. The ground state (triangle) is fixed, and both poles (lower pole in circles, higher pole in squares) get a reduction of their masses (real part), but a much stronger decrease of their widths (imaginary part), when $f_\pi$ takes $60\%$ of its vacuum value. 

As described in Ref.~\cite{Torres-Rincon:2021wyd}, the additional account for the reduction of the ground states (specially the pion mass) gives an extra decrease of all generated masses of $\Delta m \sim -100$ MeV. In particular, the lower pole of the $D_0^*(2300)$ gets bound (moving to the first Riemann surface) and becoming very close to the $D$-meson mass, while the higher-pole mass lies still $\sim 200$ MeV above them. This preliminary result---albeit rather simplified---points to a sequential degeneracy pattern in which the lower pole first becomes degenerate with the ground state, and only at higher temperatures the upper pole joins them.

\section{Conclusion} \label{sec:conclu}

In this contribution I have considered the mass degeneracy of chiral partners at high temperatures, where chiral symmetry is expected to be partially restored. I have described two effective models, the Nambu-Jona-Lasinio model for quarks and a covariant chiral EFT incorporating heavy-quark symmetry for $D$ mesons. 

In the first case, both chiral partners ($\pi$ and $\sigma$) are dynamically generated, with masses modified by the temperature. On the other hand, the chiral model contains the negative parity state as fundamental degree of freedom, while the positive parity is dynamically generated. Other options can be classified according to Table~\ref{tab:class}. In both models the generated states follow from the solution of a two-body equation: the Bethe-Salpeter equation in the NJL model, and the $T$-matrix equation in the chiral EFT.

In the NJL model the masses of the two chiral partners can be followed below and above the transition temperature and the degeneracy is clearly observed above $T_c$. In the covariant chiral EFT three different states [a double pole structure $D_{0}^*(2300)$ plus the ground state $D$] appear in the $(J,S)=(0,0)$ channel. Unfortunately this model is applicable below $T_c$ and no definite information above chiral degeneracy can be obtained in the self-consistent calculation. 

When the thermal dependence of $f_\pi$ is accounted for, it is seen that both poles move substantially in the complex energy plane becoming more bound and less massive. I showed how the lower pole approaches the ground state when $f_\pi$ is 60$\%$ of its vacuum value, while the higher pole still remains more massive. This points toward a sequential degeneracy pattern of the chiral symmetry restoration~\cite{Torres-Rincon:2021wyd}. If such pattern is supported by a more rigorous calculation, its experimental verification could be attempted in heavy-ion collisions at high energies. One would need to reconstruct the $D_0^*(2300)$ resonance into different $s-$wave decay modes, $D\pi$ and $D_s {\bar K}$, as the two poles couple more strongly to each of them (lower pole to the former, and higher pole to the latter~\cite{Montana:2020vjg}).


\paragraph{Funding information}
This research was funded by the Deutsche Forschungsgemeinschaft (German Research Foundation) grant numbers 411563442 (Hot Heavy Mesons) and 315477589 - TRR 211 (Strong-interaction matter under extreme conditions).

%
%




\bibliography{Torres-Rincon_Protvino.bib}

\nolinenumbers

\end{document}